\documentstyle[12pt,aaspp4]{article}
\begin{document}
\title{High energy gamma rays from old accreting neutron stars}
\author{P. Blasi\altaffilmark{1,2}}
\altaffiltext{1}{Dipartimento di Fisica, Universit\'a degli Studi di L'Aquila\\
Via Vetoio, 67100 Coppito (L'Aquila) - ITALY}
\altaffiltext{2}{Laboratori Nazionali del Gran Sasso\\
Statale 17 bis 67010 - Assergi (L'Aquila) - ITALY}
\begin{abstract}
We consider a magnetized neutron star with accretion from a
companion star or a gas cloud around it, as a possible source of 
gamma rays with energy between
$100$ $MeV$ and $10^{14}-10^{16}~eV$. 
The flow of the accreting plasma is terminated by a shock at the Alfv\'en 
surface. Such a shock is the site for the acceleration of 
particles up to energies of $\sim 10^{15}-10^{17}~eV$; gamma photons are 
produced in the inelastic $pp$ collisions between 
shock-accelerated particles and accreting matter. The model is applied to old
neutron stars both isolated or in binary systems. The gamma ray flux above 
$100~MeV$ is not easily detectable, but we propose that gamma
rays with very high energy could be used by Cherenkov
experiments as a possible signature of isolated old neutron stars in dense
clouds in our galaxy.
\end{abstract}


\section{Introduction}

In this paper we study a mechanism for gamma ray production in old 
magnetized neutron stars (NS's) with accretion from a dense gas cloud around 
it, or from the stellar wind of a giant companion. The NS has been supposed 
to have a surface magnetic field $B_S\sim 10^9-10^{10}~G$, and slow rotation,
so that pulsar-like activity can be neglected.\par
Old isolated NS's have already 
been extensively studied in connection with their 
possible X-ray emission by 
(Zane {\it et al.} 1995a, Zane {\it et al.} 1995)\markcite{zztt}
\markcite{ztzct}. In particular (Zane {\it et al.} 1995)
\markcite{zztt} found that inside a distance $16-30~pc$ from the Earth, about 
$10$ old isolated NS's should be present, with a soft X-ray emission in the
$0.2-2.4~keV$ band.  
A statistical analysis of their galactic 
population has been made by (Blaes and Madau 1993) \markcite{blaes_madau} 
who found that in our Galaxy there are $\sim 10^9$ 
old isolated NS's, and about $1\%$ of them are in dense clouds.\par
The accretion of gas onto magnetized compact objects has also been studied by 
many authors (see for example B\"orner 1980) in order to get a description
of the X-ray emission by them, and more specifically the interaction of the
accreting gas with the magnetic field of the NS has been investigated by 
(Arons and Lea 1976) and (Elsner and Lamb 1977) who calculated the shape
of the magnetosphere and studied the way for the plasma to reach the 
star surface. \par
(Arons and Lea 1976) claimed that a collisionless shock forms at the 
magnetosphere surface.
In the present paper we consider such a shock as the site for acceleration of 
particles up to energies $\sim 10^6-10^8~GeV$ and calculate the gamma ray 
luminosity due to $pp$ collisions in the accreting gas, in the energy range
between $100~MeV$ and $10^5-10^7~GeV$.\par  
Moreover we investigate the possibility that $300~GeV-1~TeV$ gamma rays, 
if detectable by Cherenkov experiments, could represent a
signature of old isolated NS's in our galaxy.
This paper is planned as follows: in section 2 accretion is described both on 
isolated NS's and NS's in binaries; in 
section 3 the diffusive shock acceleration 
theory is applied to calculate the luminosity in the form of high energy 
particles at the shock around the NS magnetosphere; section 4 is devoted to the 
calculation of the gamma ray flux from $pp$ collisions. Discussion and 
conclusions are presented in section 5.

\section{The accretion}
In this section we consider the accretion of matter onto a magnetized NS in two 
different scenarios: in the first one the NS is isolated and it accretes gas 
from the interstellar medium (ISM); in the 
second one the NS is in a binary system  
with a giant star, and the accretion works by the stellar wind. In both cases 
we shall study the role of the magnetic field on the accretion flow.
\subsection{Isolated Neutron Stars}
An isolated NS moving with velocity $v$ with respect to 
the circumstellar medium
having density $\rho$ can accrete it with a rate which is given by the
Hoyle-Littleton value (Hoyle and Littleton 1939)\markcite{HL}:
\begin{equation}
{\dot M}=\pi {(2GM_S)^2\over (v^2+c_S^2)^{3/2}} \rho=
3.0\times 10^{9} v_{50}^{-3} n_1 ~ g/s,
\label{eq:acc_rate}
\end{equation}
\noindent
for a flow within the accretion radius:
\begin{equation}
R_A={2GM_S \over v^2+c_S^2}=1.1\times 10^{13} v_{50}^{-2}~cm
\label{eq:acc_rad}
\end{equation}
\noindent
where $v_{50}=v/(50~km/s)$ and $n_1=n/(1cm^{-3})$ and
$c_S$ is the sound velocity in the medium. A typical value for the
velocity of isolated NS's is $v\sim 10-100~km/s$, so that they are always highly
supersonic with respect to the medium in which they move. Thus in eqs. 
(\ref{eq:acc_rate}) and (\ref{eq:acc_rad}) we neglected 
$c_S$ compared with $v$. The radius and the mass of the NS
have been taken to be $R_S=10^6~cm$ and $M_S=1~M_{\odot}$ respectively. 
If the sound velocity $c_S$ in the accreting gas remains appreciably smaller
than the free-fall velocity, we can reasonably assume, for an ideal gas, that
the accretion velocity is 
\begin{equation}
v(r)=\xi {\sqrt {2GM_S\over r}}
\label{eq:vel}
\end{equation}
\noindent
where $\xi\leq 1$ measures the small possible deviations from the free-fall
behaviour, while for the density profile it trivially follows that
\begin{equation}
\rho(r)={{\dot M}\over \pi {\sqrt {2GM_S}}} r^{-3/2}
\label{eq:dens}
\end{equation}
\noindent
holding inside the accretion radius $R_A$, and in the assumption of quasi 
spherical accretion, with a bow shock at $R_A$. In this expression for $\rho$ 
we took $\xi=1$ (Bondi and Hoyle 1944)\markcite{bondihoyle} and 
we dropped a factor $4$ due to the jump condition at the 
bow shock usually formed at the accretion radius (this shock has been 
supposed to be a strong one, so that the compression ratio is $4$).\par
Let us now introduce in this accretion scenario the large scale magnetic 
fields that are: {\it i)} the dipole-shaped magnetic field produced by the NS
\begin{equation}
B_{NS}(r)=B_S ({r\over R_S})^{-3}
\label{eq:dipole}
\end{equation}
\noindent
where $B_S$ is the surface magnetic field of the star; {\it ii)} the magnetic 
field $B_f(r)$ frozen in the accreting plasma.\par
During the inflow of the gas towards the NS, the magnetic field lines are bent 
inside the plasma; this bending causes the lines to be compressed according 
with the conservation of the magnetic flux
\begin{equation}
B_f(r)=B_f(R_A) ({r\over R_A})^{-2}
\label{eq:con_flux}
\end{equation}
\noindent
where $R_A$ is given by eq. (\ref{eq:acc_rad}). A reasonable 
assumption is that the magnetic field $B_f(R_A)$ at the boundary of the 
accretion region equals the typical ISM magnetic field 
$\sim 3\times 10^{-6}~G$. The compression of the magnetic field lines predicted 
by eq. (\ref{eq:con_flux}) is clearly limited by the reconnection rate; more 
precisely eq. (\ref{eq:con_flux}) holds up to the point where the 
hydrodynamical time scale becomes comparable with the reconnection time scale: 
$r/v_{ff}\simeq r/v_A$, where $v_A$ is the Alfv\'en velocity and
$v_{ff}$ is the free-fall one.\par
The radius at which this happens is given by
\begin{equation}
r_{eq}=7\times 10^{11}~v_{50}^{-10/3}~n_1^{-2/3}~cm.
\label{eq:r_eq}
\end{equation}
When this radius is reached reconnection compensates the increase of the 
magnetic field by compression and an equipartition value is established for 
$B_f(r)$. Thus
\begin{equation}
B_f(r) = \left \{
\matrix{B_f(R_A) (r/R_A)^{-2}& & r_{eq}<r<R_A\cr
(8\pi \rho v^2)^{1/2}& & r<r_{eq}\cr}\right.
\end{equation}
The inflow of this magnetized plasma proceeds up to the moment in which the 
energy density (pressure) of the stellar magnetic field equals that of the 
accreting gas. The radius where this condition is fulfilled is usually referred 
to as the Alfv\'en radius, and in the following we shall denote it as $R_M$, 
according with the interpretation of this radius as the boundary of the NS 
magnetosphere.\par
Two extreme cases are possible: $r_{eq}\ll R_M$ and $r_{eq}\gg R_M$.
In the first one, at the Alfv\'en radius the role of the magnetic field 
embedded in the accreting plasma can be neglected, being much less than the 
equipartition value. Thus the condition for $R_M$ is 
\begin{equation}
{B_{NS}^2(R_M) \over 8 \pi} = \rho (R_M) v^2(R_M).
\label{eq:magn_1}
\end{equation}
In the second extreme case the plasma reaches the radius $R_M$ already in 
equipartition with $B_f(r)$ and eq. (\ref{eq:magn_1}) becomes
\begin{equation}
{B_{NS}^2(R_M) \over 8 \pi} = 2 \rho (R_M) v^2(R_M).
\label{eq:magn_2}
\end{equation}
The exact equation, taking into account all the contributions is
\begin{equation}
{B_{S}^2\over 8 \pi}({R_M\over R_S})^{-6} = \rho (R_M) v^2(R_M) +
{B_f^2(R_M)\over 8 \pi}
\label{magn_compl}
\end{equation}
\noindent
where $v$ and $\rho$ are given by eqs. (\ref{eq:vel}) and (\ref{eq:dens}) 
respectively. The physical meaning of the Alfv\'en surface and 
the formation of a shock are widely discussed by 
(Arons and Lea 1976)\markcite{arons}. 
The problem of describing how the accreting matter is able to reach the surface 
of the NS is very delicate, and no self-consistent mathematical approach exists 
for taking into account the several processes involved in the interaction of 
the gas with the magnetosphere. Nevertheless some possibilities have been 
proposed: (Arons and Lea 1976) \markcite{arons} and (Elsener and Lamb 1977) 
\markcite{elsner_lamb} studied the penetration 
of the gas through the magnetosphere 
by interchange instability. (B\"orner 1980) \markcite{borner} and several other 
authors, mainly in connection 
with the problem of the X-ray emission by NS's, proposed that the plasma 
accretes down to the surface of the NS being channelled along the dipolar 
magnetic field lines, and forming on the polar caps an accretion column.

In this paper we shall study what happens at the 
shock, as the site for acceleration of protons up to high energies.

The isolated NS's which we are interested in are those located in dense 
environments ($n\sim 10^2-10^8~cm^{-3}$). Following (Blaes and Madau 1993)
\markcite{blaes_madau} in 
our galaxy there 
should be $\sim 10^9$ isolated NS, and $\sim 1\%$ of these are inside dense 
clouds. 
In the following we shall consider the cases $n=100~cm^{-3}$, $n=10^4~cm^{-3}$, 
and $n=10^8~cm^{-3}$ as far as the density of the cloud is concerned, and the 
cases $B_S=10^{9}~G$ and $B_S=10^{10}~G$ for the surface magnetic 
field of the NS. In 
table 1 the values of ${\dot M}$, $R_A$, $r_{eq}$ and $R_M$ are reported 
for $B_S=10^{10}~G$ and two velocities of the NS: $v=10~km/s$ 
and $v=80~km/s$. From these numbers 
it results that the accreting plasma reaches the magnetosphere of the NS 
already in equipartition except in the case of high velocity NS's in very large 
density clouds ($n=10^8~cm^{-3}$, $v=80~km/s$). It is easy to see that also in 
these conditions the equipartition condition holds up to $v\simeq 40~km/s$.
If equipartition is reached out of the magnetosphere the radius $R_M$ comes 
from eq. (\ref{eq:magn_2}):
\begin{equation}
R_M=1.5\times 10^9~v_{50}^{6/7} n_1^{-2/7} B_{10}^{4/7}~cm
\label{eq:r_m}
\end{equation}
\noindent
where $B_{10}=B_S/10^{10}G$.\par
\placetable{table_1}
In these calculations we neglected all the effects coming from the rotation of 
the NS. This is equivalent to require that
(see Treves {\it et al.} 1995)\markcite{tcl}: 1) the relativistic wind produced 
beyond the light cylinder is not able to stop the accretion at the radius 
$R_A$; 2) the accretion flow velocity at $R_M$ is larger than the corotation 
velocity. \par
Condition 1) means, for the period $P$ of the NS, that \par
$$P>1.1~B_{10}^{1/2}~({\dot M}/10^{11}g/s)^{-1/4}~(R_A/10^{14}cm)^{1/8}~s,$$
\par\noindent  
while condition 2) gives \par
$$P>4.9({\dot M}/10^{11}g/s)^{-3/7}~s.$$\par 
It is easy to check that these conditions are fulfilled by old isolated NS's, 
for which the rotation period is usually of a few seconds.
\subsection{Neutron Stars in binaries}
We consider here a specific model for a close binary system with a NS and a 
giant with a stellar wind. The velocity of the wind far from the star is 
$v_W=10^6-10^8~cm/s$; the rate of mass loss will be denoted by 
${\dot M_{loss}}$. The density of wind matter at distance $r$ from the giant is
\begin{equation}
\rho_W(r)={{\dot M_{loss}}\over 4\pi v_W r^2}.
\label{eq:wind}
\end{equation}
This radial outflow is appreciably influenced by the presence of the NS at a 
distance from it equal to the accretion radius
\begin{equation}
R_A={2GM_S\over v_W^2}=2.7\times 10^{12}~({v_W\over 10^7cm/s})^{-2}
\label{eq:w_acc}
\end{equation}
\noindent
where $M_S$, as usual, is the NS mass. We shall use here a simplified model, 
similar to that used in (Berezinsky {\it et al.} 1996) \markcite{bbh} 
(the accreting compact object was a white dwarf there) 
where at $\sim R_A$ a bow shock forms, and inside the shock the 
accretion onto the NS 
becomes spherically symmetric, with a density given again by 
eq. (\ref{eq:dens}), where ${\dot M}$ is now connected to ${\dot M_{loss}}$ by 
geometrical considerations. In particular if $d$ is the interbinary 
distance, we can write
\begin{equation}
{\dot M}\simeq {\dot M_{loss}} ({R_A\over d})^2=7.3\times 10^{-8} d_{13}^{-2}
({v_W\over 10^7cm/s})^{-4}~({{\dot M_{loss}}\over 10^{-6} M_{\odot}/yr})
~M_{\odot}/yr,
\label{eq:rate}
\end{equation}
where $d_{13}=d/(10^{13}cm)$. In the following we shall use 
${\dot M_{loss}}=10^{-6}~M_{\odot}/yr$.
The basic features of the accretion and of the 
interaction between the accreting plasma and the stellar magnetic field are 
the same as those previously explained for isolated NS's. 
However in the case of 
binaries it is more difficult to fix the boundary conditions on the magnetic 
field at the accretion radius, and some {\it ad hoc} assumptions about the 
stellar magnetic field of the companion should be required. For simplicity we 
shall assume that during accretion the equipartition is reached due to the fact 
that the reconnection time scale becomes comparable with the 
hydrodynamical time scale before the Alfv\'en radius is reached. Thus 
eq. ({\ref{eq:magn_2}) holds, and we can write:
\begin{equation}
R_M=6.2\times 10^6 ~ B_{10}^{4/7} {\dot M_{-8}}^{-2/7}~ cm
\label{eq:rmbin}
\end{equation}
\noindent
where ${\dot M_{-8}}$ is the accretion rate in units of $10^{-8}~M_{\odot}/yr$.
\par
In this paper we are interested in NS's that do not show pulsar behaviour, so 
that, in the case of binaries too, we assume that the NS rotates slowly and 
that its magnetic field is not larger than $10^{10}~G$.\par
The discussion made in the case of isolated NS's about the formation of a shock 
as a consequence of the interaction of the accreting plasma with the magnetic 
wall at the boundary of the magnetosphere holds in this case as well; in the 
case of binaries it is possible that the geometry of the accretion is 
modified with respect to the spherical 
one, with the formation of an accretion disk. 
This happens when the giant fills its Roche Lobe, and the transfer of matter to 
the NS works by the inner Lagrangian point instead than by the stellar wind. 
In this situation the simplified model used here is no longer valid and the 
effect of the magnetic field should be the disruption of the 
internal part of the disk itself. We shall not consider this case here. 
\section{The acceleration}
The shock acceleration mechanism is discussed in several reviews 
(Jones and Ellison 1991, Blandford and Eichler 1987, Drury 1983) 
\markcite{jones} \markcite{be} \markcite{drury} and 
we shall 
not stress here the technical details. In this paper we propose that the shock 
which is formed on the boundary of the magnetosphere of a NS, according to the 
mechanism described in the previous sections, can accelerate some fraction of 
the accreting particles up to very high energies, and that the reinteraction of 
the accelerated particles with the gas produces a gamma ray signal.\par
The shock is located at the radius $R_M$ defined by eqs. ({\ref{eq:r_m}) and 
(\ref{eq:rmbin}) for the case of isolated NS's and NS's in 
binaries respectively. In 
the following we shall discuss the two cases separately. \par
It is easy to estimate the luminosity in the form of accelerated 
particles (we assume they 
are protons) if we introduce an acceleration efficiency  $\eta$:
\begin{equation}
L_{acc}=\eta~{G M {\dot M}\over R_M}.
\label{eq:l_acc}
\end{equation}
The value usually used for $\eta$ is $\sim 0.1$ (Berezinsky {\it et al.} 1990)
\markcite{betal}. In 
the case of isolated NS's, by 
using eqs. (\ref{eq:acc_rate}) and (\ref{eq:r_m}) we have
\begin{equation}
L_{acc}^{is}=2.6\times 10^{26}~\eta~v_{50}^{-27/7}~B_{10}^{-4/7}~n_1^{9/7}~
erg/s.
\label{eq:l_is}
\end{equation}
For NS's in binaries, by eq. (\ref{eq:rmbin}):
\begin{equation}
L_{acc}^{bin}=1.4\times 10^{37}~\eta~{\dot M_{-8}}^{9/7}~B_{10}^{-4/7}~erg/s.
\label{eq:l_bin}
\end{equation}
The maximum energy reachable by this acceleration mechanism can be calculated 
by the condition that the diffusion time of the accelerated particles becomes 
equal to the typical loss time, mainly due, in this case, to inelastic $pp$ 
collisions, whose typical cross section is $\sigma_0=3.2\times 10^{-26}~cm^2$.
\par
The diffusion time is given by
\begin{equation}
t_{diff}={3\over u_1 - u_2}~({D_1\over u_1} + {D_2\over u_2})
\label{eq:t_diff}
\end{equation}
\noindent
where the subscript $2$ refers to the region $r<R_M$ and the subscript $1$ 
refers to $r>R_M$. $D_i=E(eV) c / 300 B_i(Gauss)$ are the diffusion 
coefficients (we adopt here the Bohm's values), and $u_i$ are the flow 
velocities on the two sides of the shock. In the assumption of a strong shock
we have $u_1/u_2=4$ and eq. (\ref{eq:t_diff}) becomes
\begin{equation}
t_{diff}={4+\sqrt{2}\over 75} {E(eV)\over B_S} {R_M^4 c\over 2 G M_S R_S^3} 
\label{eq:t_diff_1}
\end{equation}
\noindent
where we used eq. (\ref{eq:dipole}) for $B_2=B_{NS}(R_M)$ and we used the 
equipartition condition in the accreting plasma.\par
The typical loss time is
\begin{equation}
t_{loss}={1\over n(R_M) \sigma_0 c} = 
{\pi \sqrt{2 G M_S} m_p \over {\dot M} \sigma_0 c} R_M^{3/2}.
\label{eq:loss}
\end{equation}
The equation for the maximum energy, obtained equating $t_{diff}$ and 
$t_{loss}$, is thus:
\begin{equation}
E_{max}(eV)=1.1\times 10^{40} {R_M^{-5/2} B_S\over {\dot M}}.
\label{eq:emax}
\end{equation}
This gives the results shown in table 2 for isolated NS's (except the case 
shown in the last column where the gas reaches the magnetosphere not in 
equipartition) and
\begin{equation}
E_{max}^{bin}(eV)=1.8\times 10^{15} ~ B_{10}^{-3/7} ~ {\dot M_{-8}}^{-2/7}
\label{eq:emax_bin}
\end{equation}
\noindent 
for NS's in binaries.\par
The differential spectrum of the accelerated particles is taken as 
${\dot N_p(E)}=K(E+E_0)^{-\gamma}$, where the constant $K$ is easily obtained 
by the condition
\begin{equation}
\int_{E_{th}}^{E_{max}} dE ~ E ~ {\dot N_p(E)}~=~L_{acc},
\end{equation}
\noindent
where $E_0\sim 1~GeV$ and $E_{th}$ is the threshold energy for pion production
in $pp$ collisions. 
The exponent $\gamma$ in the shock acceleration theory 
is connected to the compression ratio $r=u_1/u_2$ by the expression 
$\gamma=(r+2)/(r-1)$. For a strong shock $r=4$ and $\gamma=2$.\par
From the previous discussion it results that accreting 
old NS, both isolated or in 
binary systems, can accelerate particles up to $\sim 10^{15}-10^{17}~eV$, by 
the shock arising on the boundary of the NS magnetosphere. The 
luminosity in the form of accelerated particles and reference velocity 
$50~km/s$ is shown in table 2.

\section{Gamma ray production}
Gamma rays are produced by $pp$ inelastic collisions between the 
accelerated particles and the accreting gas. The main mechanism  for gamma ray 
production is the decay of the neutral pions generated by the reaction 
$p+p\to \pi^0+X$.\par
This process has been studied in detail by many authors, and the relevant 
cross sections, taking into account pion multiplicities, can be found in 
(Dermer 1986) \markcite{dermer}.
\par
The total number of gamma rays per unit time can be written as follows:
\begin{equation}
Q_{\gamma}=2~{x\over m_p}~\int_{E_{th}}^{E_{max}} dE~{\dot N_p(E)}~
<\xi\sigma(E)>
\label{eq:number}
\end{equation}
\noindent
where $x$ is the grammage out of the magnetosphere and $<\xi\sigma>$ is 
a multiplicity weighted cross section for the reaction $p+p\to \pi^0+X$ 
(Dermer 1986) \markcite{dermer}.

The real value for $x$ 
could be appreciably larger than the integral of $\rho$ onto a straight line 
out to some maximum distance (of the order of the accretion radius), due to the 
presence of the magnetic field, which is responsable for the curvature of the 
particle trajectories and therefore of a larger column density suffered 
by these particles.\par
The integral in eq. (\ref{eq:number}), if $x$ is simply calculated 
by integrating over straight radial lines, is reported in table 2 for isolated 
NS's, for several values of the gas density $n$ and of the surface magnetic 
field. For NS's in binary systems (we took ${\dot M}=10^{-8}~M_{\odot}/yr$, 
$d=10^{13}~cm$) the calculation gives
\begin{equation}
Q_{\gamma}^{bin}\simeq 4\times 10^{36}~B_{10}^{-6/7}~s^{-1}
\label{eq:qbin}
\end{equation}
\noindent
It is clear from table 2 that for isolated NS's with $v=50~km/s$ and for 
typical distance scale of $\sim 100~pc$, no gamma ray signal can be detected in 
the region $E>100~MeV$ for $n=100-10^4~cm^{-3}$, which is the usual range of 
average density in nearby dense clouds. From this point of view giant molecular 
clouds (GMC) are of particular interest: the density in 
their cores can reach the 
value $10^8~cm^{-3}$ (Turner 1988)\markcite{Turner} so that the gamma ray 
flux above $100~MeV$ 
becomes detectable up to a distance $\sim 800~pc$. In this range of distances 
many giant molecular clouds are present (see Dame {\it et al} 1987 
\markcite{dame}), the closest of which is Taurus-Auriga, at $\sim 140~pc$. From 
table 2 we see that an isolated NS in the core of a GMC at such a distance 
should give a signal of $10^{-6}~photons~cm^{-2}~s^{-1}$ above $100~MeV$. In
(Blaes and Madau 1993)\markcite{blaes_madau} 19 of such clouds are listed with 
the correspondent expected number of NS's. All of them are inside a distance of 
$830~pc$ from the Earth and isolated NS's in their possibly very dense cores 
could produce a detectable gamma ray signal with $E>100~MeV$.\par
In table 2 it is also reported the flux of gamma rays above $300~GeV$ that 
could be detected by a Cherenkov imaging experiment such as Whipple, whose 
sensitivity is $\sim 10^{-13}~cm^{-2}~s^{-1}$.\par
\placetable{table_2}
While the detectability on scales of $50-100~pc$ from isolated NS's in a cloud 
with $n=10^4~cm^{-3}$ is limited to low velocity NS's ($v\sim10~km/s$), the 
signal with $E>300~GeV$ from the very dense cores of the GMC is possible on 
galactic scales ($d\sim 9-10~kpc$).\par
The calculation of the high energy flux has been carried out by introducing the 
photon yield $Y_{\gamma}$:
\begin{equation}
Q_{\gamma}(E>0.3TeV)\simeq {\sigma_0\over m_p}~x~Y_{\gamma}
\int_{0.3TeV}^{E_{max}} dE~{\dot N_p(E)},
\end{equation}
\noindent
where $x$ is the grammage and $Y_{\gamma}\sim 0.1$; the photon yield 
(see Berezinsky {\it et al.} 1990) \markcite{betal}
represents the number of photons with energy $E$ produced by one 
proton with the
same energy which undergoes a $pp$ collision. 

\section{Discussion and conclusions}
We studied the accretion of matter onto compact objects with surface magnetic 
field $B_S=10^{10}~G$ and with rotation slow enough to allow accretion and not 
to show pulsar-like activity.\par
Such objects could be old NS's, for which it has been foreseen a 
decrease in the surface magnetic field down to $10^9-10^{10}~G$
 (Phinney and Kulkarni 1994) \markcite{pk}.\par
As a result of the interaction between the magnetic field of the star and the 
accreting plasma, a magnetosphere is formed around the NS, bounded by a 
collisionless shock (Arons and Lea 1976)\markcite{arons}, where 
the pressure of the gas equals that due to the 
magnetic field of the NS. Even if the existence of such a shock has been 
assumed by several authors, no conclusive agreement has been reached 
about it, due to the fact that a self-consistent mathematical description of 
the accretion in presence of a strong magnetic field does not exist. In this 
paper we assumed the existence of the shock and proposed that it could 
accelerate nuclei (protons) up to high energies (see also Shemi 1995)
\markcite{shemi}, according with the usual 
mechanism of the diffusive shock acceleration.\par
These accelerated particles can produce a gamma ray signal due to the inelastic 
$p+p\to \pi^0+X$ collisions with $\pi^0$ decay in $\gamma\gamma$, where the 
target is provided by the accreting gas.\par
The luminosity in the form of accelerated particles from isolated NS's strongly 
depends on two parameters, the 
matter density around the NS and the velocity of the NS. 
The NS's we considered here  are those contained in dense clouds (there are 
$10^7$ of such objects in the Galaxy) where the gas density ranges between 
$100~cm^{-3}$ and $10^8~cm^{-3}$ (in the cores of the giant molecular clouds).
A statistical study of the velocity distribution of these NS's has been made by 
Blaes and Madau (1993) \markcite{blaes_madau} who found that 
$\sim 6\%$ of these objects have $v<20~km/s$, 
$\sim 22\%$ have $v<40~km/s$ and $\sim 50\%$ have $v<72~km/s$. \par
In table 1 we shown the accretion parameters for $n=100,~10^4,~10^8~cm^{-3}$ 
and for two extreme velocities, $10~km/s$ and $80~km/s$, verifying that for old 
NS's the role of the rotation can be neglected.\par
In table 2 we shown the acceleration data and the fluences of gamma rays with 
$E>100~MeV$, fixing the velocity at $50~km/s$ and distinguishing the cases $B_S=
10^9~G$ and $B_S=10^{10}~G$, for the usual range of densities. In particular we 
also estimated the flux of very high energy gamma rays ($E>300~GeV$) detectable 
by the Cherenkov experiments such as Whipple. From these data it seems 
hopeless to observe with present detectors the $100~MeV$ radiation 
produced by the $pp$ collisions in 
gas clouds with $n\sim 100-10^4~cm^{-3}$, at least for distance scale of $\sim 
100~pc$, unless for very low velocity NS's ($v\leq 10~km/s$) which are however 
a very small percentage of the total. \par
Much more interesting is the situation of the cores of the GMC
($n\sim 10^8~cm^{-3}$): in this case the $100~MeV$ radiation is 
detectable on distances of $800~pc$ and the very high energy part of the gamma 
ray spectrum ($E>300~GeV$) could be detected by Whipple-like experiments on 
galactic scales ($9-10~kpc$).
Many GMC are located within such distances: the closest is Taurus-Auriga, at 
$140~pc$, which could contain up to $30$ NS's; about $300$ NS's should be 
present in Cignus rift, $700~pc$ far away, and many other GMC can be found 
within $800-900~pc$ (Blaes and Madau 1993, Dame {\it et al.} 1987)
\markcite{blaes_madau}\markcite{dame}.\par
Isolated NS's in the dense cores of GMC would necessarily be bright X ray 
emitters, because the matter stopped at the magnetosphere wall is later 
channelled to the polar caps, liberating a luminosity which is two-four orders 
of magnitude larger than the gamma ray one. However such X rays will hardly 
leave the dense region, due to the strong Compton absorption. Besides accretion 
of gas down to the star surface could be stopped, for some values 
of the parameters, due to the fact that low velocity NS's in very large density 
environments should accrete at super-Eddington rate: radiation pressure could 
introduce in this case some non stationary accretion regime.
Thus gamma rays would be a unique tool for the study of the isolated NS's 
in overdense regions. On the other hand X rays in the $0.2~keV$ region 
remain the most powerful way to 
observe isolated NS's in moderately dense clouds, where the gamma ray flux 
should be too small.\par
Here we discussed also the case of NS's not showing pulsar-like activity, in 
binary systems, but the conclusions reached about them are strongly parameter 
and model dependent. This is a consequence of two factors: first of all the 
accretion rate onto the NS varies abruptly when the geometry of the binary is 
changed (e.g. interbinary distance, size of the companion), and 
therefore the position of the magnetic wall is also changed, and with it the 
luminosity in the form of cosmic rays. The second 
factor is that the geometry of the 
accretion can probably become far from being spherical (disk accretion); in the 
case of disk accretion, the effect of the stellar magnetic field is usually to 
disrupt the part of the disk inside the Alfv\'en radius, and no shock is 
probably produced in this case, or at least it should be not useful for 
accelerating particles. Nevertheless, in the cases in which our calculation 
holds, the signal from old NS's in binaries 
can be detected by EGRET if the source is located inside $\sim 
500~pc$ from the Earth. \par

\acknowledgments{The author thanks V.S. Berezinsky and B.I. Hnatyk for many 
useful discussions; in particular he is grateful to V.S. Berezinsky for 
having encouraged this calculation.}

\newpage
\begin{table}
\caption{The values of the parameters for the accretion onto
an isolated neutron star for $B_S=10^{10}~G$. \label{table_1}}
\begin{center}
\begin{tabular}{l c c c c c c}
\hline \hline
 & \multicolumn{2}{c}{$100~cm^{-3}$} &
\multicolumn{2}{c} {$10^4~cm^{-3}$} & \multicolumn{2}{c} {$10^8~cm^{-3}$} \\

parameters & $10~km/s$ & $80~km/s$ & $10~km/s$ & $80~km/s$ & $10~km/s$ 
& $80~km/s$ \\ \hline

$R_A$ (cm) & $2.7\times 10^{14}$ & $4.3\times 10^{12}$ & $2.7\times 10^{14}$ & 
$4.3\times 10^{12}$ & $2.7\times 10^{14}$ & $4.3\times 10^{12}$ \\

${\dot M}$ (g/s) & $3.7\times 10^{13}$ & $7.3\times 10^{10}$ & 
$3.7\times 10^{15}$ & 
$7.3\times 10^{12}$ & $3.7\times 10^{19}$ & $7.3\times 10^{16}$ \\

$r_{eq}$ (cm) & $6.9\times 10^{12}$ & $6.7\times 10^{9}$ & $3.2\times 10^{11}$
 & $3.1\times 10^{8}$ & $6.9\times 10^{8}$ & $6.8\times 10^{5}$ \\

$R_M$ (cm) & $1.0\times 10^{8}$ & $6.0\times 10^{8}$ & $2.7\times 10^{7}$ & 
$1.6\times 10^{8}$ & $2.3\times 10^{6}$ & $1.4\times 10^{7}$ \\ \hline

\end{tabular}
\end{center}

\caption{The values of the parameters of the acceleration and the gamma ray 
production by $pp$ collisions. \label{table_2}}
\begin{tabular}{l c c c c c c}
\hline \hline
 & \multicolumn{2}{c}{$100~cm^{-3}$} &
\multicolumn{2}{c} {$10^4~cm^{-3}$} & \multicolumn{2}{c} {$10^8~cm^{-3}$} \\

parameters & $10^9~G$ & $10^{10}~G$ & $10^9~G$ & $10^{10}~G$ & $10^9~G$ 
& $10^{10}~G$ \\ \hline

$E_{max}$ (eV) & $3.0\times 10^{17}$ & $1.1\times 10^{17}$ & 
$8.1\times 10^{16}$ & 
$3.0\times 10^{16}$ & $6.0\times 10^{15}$ & $1.6\times 10^{15}$ \\

$L_{acc}$ (erg/s) & $3.6\times 10^{28}$ & $1.0\times 10^{28}$ & 
$1.3\times 10^{31}$ & 
$3.6\times 10^{30}$ & $2.0\times 10^{36}$ & $5.0\times 10^{35}$ \\

$x$ ($g/cm^2$) & $1.1\times 10^{-6}$ & $5.6\times 10^{-7}$ & $2.1\times 10^{-4}$
 & $1.1\times 10^{-4}$ & $7.8$ & $4.0$ \\

$Q_{\gamma}(>0.1~GeV)$ ($s^{-1}$) & $8.0\times 10^{21}$ & 
$1.2\times 10^{21}$ & $6.0\times 10^{26}$ & 
$9.0\times 10^{25}$ & $7.9\times 10^{36}$ & $1.1\times 10^{36}$ \\ 

$Q_{\gamma}(>0.3~TeV)$ ($s^{-1}$) & $8.4\times 10^{18}$ & 
$1.2\times 10^{18}$ & $6.2\times 10^{23}$ & 
$9.7\times 10^{22}$ & $4.2\times 10^{33}$ & $5.8\times 10^{32}$ \\ \hline 

\end{tabular}

\end{table}

\end{document}